\documentclass[aps,pre,onecolumn,groupedaddress,showpacs]{revtex4}

\begin{document}
 

\title{Competing tunneling trajectories in a 2D potential with variable topology as a model for
quantum bifurcations. }
 
\author{V.A.Benderskii} 
\affiliation {Institute of Problems of Chemical Physics, RAS \\ 142432 Moscow
Region, Chernogolovka, Russia} 
 
\author{E.V.Vetoshkin} 
\affiliation {Institute of Problems of Chemical Physics, RAS \\ 142432 Moscow
Region, Chernogolovka, Russia}
 
\author{E. I. Kats} 

\affiliation{Laue-Langevin Institute, F-38042, Grenoble, France,
and L. D. Landau Institute for Theoretical Physics, RAS, Moscow, Russia}

\author{H.P. Trommsdorff}
\affiliation{
Laboratoire de Spectrom'trie Physique, Universit' Joseph-Fourier de Grenoble,
CNRS (UMR 5588), B.P. 87, F-38402 St. Martin d'HŠres Cedex, France}

\date{\today}
 
\begin{abstract}                                                                                                
We present a path - integral approach to treat a 2D model of a quantum bifurcation. 
The model potential has two equivalent minima separated by one or two saddle points, 
depending on the value of a continuous parameter. 
Tunneling is therefore realized either along one trajectory or along two equivalent paths. 
Zero point fluctuations smear out the sharp transition between these two 
regimes and lead to a certain crossover behavior. 
When the two saddle points are inequivalent one can also have a first 
order transition related to the fact that one of the two trajectories 
becomes unstable. We illustrate these results by numerical 
investigations. Even though a specific model is investigated here, 
the approach is quite general and has potential applicability 
for various systems in physics and chemistry exhibiting multi-stability and tunneling phenomena.
\end{abstract}
 
 
\maketitle
 
\section{Introduction}
\label{intr}                                                                                                    

Molecules possessing more than one stable or metastable configurations 
(so-called non-rigid molecules 
\cite{be1}, \cite{be2})  
are interesting as many of their properties differ considerably from those 
of rigid molecules. In addition, such molecules have practical 
applications as building blocks of present and future display 
and sensor technologies. Theoretical modelling of non-rigid molecules 
is often hampered by a lack of detailed knowledge concerning 
their eigenstates and eigenfunctions.
As a rule, vibrational spectra of the non-rigid molecules 
are characterized by local oscillations around the minima 
and tunneling splittings due to transitions between 
these quasi-local states. When only one ''coordinate'' (the reaction path) 
is related to a large amplitude motion while all other degrees of 
freedom can be approximated as small amplitude oscillations 
\cite{be3}, \cite{be4}, 
the behavior of the system is determined completely 
by a single minimum action trajectory. The dynamics 
becomes more complex when more than one coupled large 
amplitude motions (i.e. more than one strongly fluctuating 
variables) are present. In this case, several minimum energy paths 
connect minima of the potential energy surface (PES). 
Phenomena like instabilities or bifurcations can appear in 
certain regions of parameters describing the PES, and the analysis 
of such a situation is the main issue of the present paper.

In the general case of a PES with n minima, states of the system 
can be described using $n$ sets of eigenfunctions, quasi-localized 
near each minimum, while tunneling
between different states is related to 
the overlap between these functions. The characteristic 
energy scale for tunneling, usually much smaller than the thermal energy, 
is determined by the chemical nature of the non-rigid molecule, 
or, in other words, by the quantum chemistry of the system. 
Depending on the parameters describing the PES of a non-rigid 
molecule, one may expect different types of behavior, which can 
be treated in terms of quantum instabilities also called quantum phase transitions.

Quantum phase transitions (that occur in a quantum mechanical system 
at zero temperature as a function of some non-thermal control parameter) 
and related phenomena, like quantum bifurcations and instabilities 
have been a subject of great theoretical interest in recent years 
(see e.g. monograph \cite{SA99} and references herein). 
Unlike ordinary phase transitions, quantum transitions occur between
ground states and involve negligible changes in entropy.
Examples include transitions in Quantum Hall systems, 
localization phenomena, and the superconductor - insulator transition 
in two-dimensional systems 
\cite{SA99}, \cite{SG97}.
Usually these phenomena take place in many particle 
systems driven by competing interactions (a well known example 
is the superconductor - normal metal quantum phase transition at $T = 0$ 
arising from electron - electron
correlation competing with electron - phonon interaction  \cite{MA62}). 
As a result of the competition, a quantum bifurcation may 
occur at a critical value of the coupling constant, leading 
to singularities of some properties. The most robust features 
of quantum bifurcations are divergences or singularities in 
certain characteristics of the system at the critical value of 
the coupling constant and drastic changes of the 
behavior below and above. Both classical and quantum critical 
points are governed by a divergent correlation length, 
although quantum systems possess additional properties that 
do not have classical counterparts. The fact that the ground state wavefunction 
undergoes a qualitative change at a quantum instability is one of these and is the subject of this paper.

Remarkably, similar features may arise in few-body systems. 
A well-known example is the so-called level crossing phenomenon 
\cite{LL65} when an excited level becomes the ground state 
at a critical value of the interaction
parameter. In this paper we analyze another type of quantum instability 
taking place for a 2D PES changing its topology as 
a function of a continuous parameter $\alpha $ and illustrate the 
results by numerical investigations. The model PES 
has two equivalent minima separated by one or two saddle points. 
The change from one to two saddle points takes place for 
a certain critical value of  $\alpha = \alpha _c$.   
In addition, the PES may be symmetric or asymmetric with 
respect to the second coordinate, perpendicular to the one 
relating the two minima. For the symmetric case (two 
equivalent saddle points) there are two regions 
(I and II) of the phase diagram, where tunneling 
is realized either along one minimum action trajectory 
essential for the semiclassical description of the motion (region I), 
or along two equivalent paths (region II). Zero point 
fluctuations smear out a sharp second order transition between 
the two regions and lead to a certain crossover behavior. 
For the asymmetric case, one can have also a first 
order phase transition related to the fact that one of the 
two trajectories becomes unstable.

The competition between trajectories in this model plays 
the role of competing interactions in many body systems with 
quantum bifurcations. Divergences in thermodynamic 
quantities are mapped onto a certain singular behavior 
in our model (critical fluctuations at finite-temperatures 
have their equivalent in the zero-point quantum delocalization) 
and, in fact, reflect more fundamental phenomena. 
Loosely speaking, any continuous quantum bifurcation is related 
to specific conditions (or specific values of parameters) when 
the lowest excitations become gapless and a qualitative 
change in the nature of the frequency spectrum occurs.

In the literature \cite{KS00}, \cite{SK01}, 
several analogies have been proposed to describe the 
ground state stability of a small system (such as an 
atom or a molecule) in terms of phase transitions, 
mapping a $d$-dimensional quantum systems onto a $d+1$-dimensional classical system. 
This formal
equivalence (first proposed long ago \cite{jo1}, \cite{jo2})
is based on the observation that the quantum action 
includes not only an integration over the space variables but 
includes also one imaginary time dimension, which, for $T = 0$, 
is of infinite extent. The quantum -
classical mapping is a very general fact. One can always 
reinterpret the imaginary time functional integral of a $d$-dimensional 
quantum theory as a finite
temperature Gibbs distribution function of a $d+1$-dimensional classical theory.

Naively, this seems to imply that quantum bifurcations 
are not very interesting. However, as elsewhere in science, 
the devil is in the details of this mapping. 
For example it turns out 
\cite{SA99} 
that there is no 
guarantee that the Gibbs weights, found by such a mapping, 
are positive (they can even be complex valued). This implies 
that one may not use the mapping blindly, i.e. for any 
arbitrary system. Thus a direct and explicit treatment 
of the quantum models is required.

Instead of the divergence of the correlation length (which is the 
cornerstone of traditional descriptions of criticality 
in determining the scaling behavior of all other quantities) 
any quantity that changes its scaling behavior at $\alpha = \alpha _c$ 
can be used. Thus, it is not
necessary to
restrict analogies of phase transition to the search for 
the points where the correlation length diverges. 
As shown below, in the present model the wave functions itself 
manifest a characteristic critical behavior.

The remainder of this paper is organized as follows. 
In Section II we investigate a model 2D potential having 
two equivalent minima, and two saddle points. The shape and the topology of 
the potential depend on one continuous parameter, $\alpha $, 
playing the role of a controlling parameter
in a
phase transition terminology. At the critical value $\alpha = \alpha _c$, two saddle points (and
the maximum of the
potential in between) merge into one saddle point. 
In this Section we also present a qualitative picture 
of the quantum bifurcation. Section III is devoted 
to the numerical verification of the qualitative scenario. 
Minimum energy and extremal tunneling trajectories of the potential 
are determined. In the Section IV the results presented in 
the Section III are generalized to an anisotropically deformed PES, 
with the aim to relate predictions of the theoretical results 
to possible experimental verifications such as the effect 
of partial deuteration. Some specific examples of 
non-rigid molecules are also discussed in this section. 
Finally, in Section V we conclude.

\section{Instanton approach to a model 2D potential.}
\label{instanton}

In order to investigate the scenario of quantum bifurcations described above, 
one should solve the Schr\"odinger equation for a 2D PES of variable shape 
and topology. No standard
method
of solution is known for a 2D PES of general shape. 
In the semiclassical approximation (the concern of this paper), 
the commonly used WKB method 
\cite{LL65}, \cite{HE62},
is reduced to matching wave functions between classically allowed and forbidden regions. 
This matching is easily achieved in the 1D case with only one 
strongly fluctuating variable. Technically, already for two 
dimensions with one strongly and one weakly fluctuating 
variable the procedure becomes tedious and computationally demanding. 
However for higher dimensional PES and for more than coupled 
large amplitude motions, fundamental difficulties are encountered of 
how to match many-valued semiclassical wave functions at a non-trivial set 
of boundaries between allowed and forbidden regions in multidimensional 
phase space. To the best of our knowledge this problem has not been 
investigated previously in any general form.

However, there is an alternative to the universal WKB semiclassical 
formalism, the so-called extreme tunneling trajectory 
or instanton method, which is not only very efficient 
for calculating globally uniform wave functions of the 
ground state 
(as initially formulated 
\cite{PO77}, \cite{CO}) 
but which can also be adapted for the description of low-energy excited 
states \cite{BM}, \cite{BV99}, \cite{BV00}  
(and which can even be extended with reasonable accuracy to highly excited states 
\cite{BK02}).                                                                                    
The instanton method eliminates much of the tedious WKB calculations 
because there are no classically allowed regions in the 
formalism (and as a consequence there are no singularities of semiclassical 
wave functions). Within the instanton method, bifurcations or 
instabilities are simply related to the existence of 
multiple-valued solutions of the equations of motion, i.e. 
appearance of two (or more) solutions with similar values 
of their action. Unlike WKB wave functions, 
which have singularities at the boundaries between classically allowed and 
forbidden regions, the variations over the space 
of instanton wave functions can be followed continuously. 
The price to pay for this advantage is related to the nonexistence of 
classically allowed regions and the absence of natural 
Bohr - Sommerfeld quantization rules. 
In the instanton approach, these rules must be established 
by different methods. In addition, the instanton equations 
contain second order turning points whereas the WKB equations 
have only linear, first order, turning points.

The main steps of the instanton method, used to determine semiclassical 
solutions to the Schr\"odinger equation in the potential $V(X,Y)$, are:
\begin{itemize} 
\item i) 
The so-called Wick rotation in phase space, corresponding to the transformation 
to imaginary time ($t \to - i t$), is made. 
\item ii) 
Solutions of the Schr\"odinger equation are
represented in the form 
\begin{eqnarray} 
\label{b1} \Psi = A(X , Y) \exp (-\gamma W (X , Y)) \, ,
\end{eqnarray}                                                                                                  
$\gamma \gg 1$ is a large semiclassical parameter, 
$\gamma \equiv m \Omega _0 a_0^2/\hbar $,  
where $m$ is a mass of a particle, $a_0$ is a ''microscopic'' characteristic 
length of the problem,
e.g. the
tunneling distance, $\Omega _0$ is a characteristic frequency, 
e.g. the oscillation frequency around the
potential
minimum, $\hbar $ is set equal to 1, 
measuring energies in frequency units, and $W$ has a meaning of action. 
\item iii) 
The terms in the Schr\"odinger equation, first and second order 
in $\gamma ^{-1}$, become identically zero, if
the
wave function in the form of Eq. (\ref{b1}) 
satisfies the time independent Hamilton-Jacobi equation for 
$W$ and the so-called transport equation for $W$ and $A$ 
(see below and also \cite{BM}, \cite{BV99},
\cite{BV00} for details).
\end{itemize} 

Below, the instanton approach is used to 
determine semiclassical solutions of the Schr\"odinger 
equation in the following model 2D PES: 
\begin{eqnarray} 
\label{b2} 
V(X , Y) = \frac{1}{2}(1-X^2)^2 + \frac{\omega ^2}{1-\alpha }\left
[\frac{1}{2} (X^2 - \alpha )Y^2 + \frac{1}{4} Y^4 \right ] \, .
\end{eqnarray} 
The PES given by Eq. 
(\ref{b2}) is dimensionless: coordinates 
$X$ and $Y$ are measured in units of the distance, $a_0$, 
between two minima along $X$, the energy is
measured in
units of the oscillation frequency, $\Omega _0$, in one of the wells. 
The prefactor of the second term of
Eq. (\ref{b2}) is chosen such that the oscillation frequency along $Y$ 
(at $X = \pm 1$) is independent of $\alpha $,
and equals the parameter $\omega $. 
The potential, Eq. (\ref{b2}), is bound for $\alpha < 1$ and has in this case the
following stationary points: 
$\{ X = \pm 1\, , \,  Y = 0\} $ are two equivalent minima; 
$\{X = 0\, , \, Y = 0\}$ is a saddle point for $\alpha < 0$ and
becomes a maximum at $0< \alpha < 1$; $\{X = 0\, , \, Y = \pm \sqrt \alpha \}$ 
are two saddle points
at $0 < \alpha < 2/(2 + \omega ^2)$, becoming
minima when $\alpha  > 2/(2 + \omega ^2)$ with the appearance of 
four additional saddle points. In the
following, the discussion is restricted to the region of $0 < \alpha < 2/(2 +\omega ^2)$, 
where additional
stationary
points, having no ''critical'' behavior, are absent. 
The bifurcation phase diagram in the plane of the parameters $\alpha \, , \, \omega ^2$ 
is shown in
Fig. 1. For a classical
system, one would have a continuous bifurcation 
at the line $\alpha  = 0$ from a behavior governed by one minimum 
energy trajectory (region I in the
Fig. 1) to
dynamics with two minimum energy paths passing through 
the two saddle points (region II). 
Representative equipotential maps of PES corresponding to the 
two regions are shown in Fig. 2. 
As discussed in the next Section, quantum zero point 
fluctuations shift and deform the transition line and 
smear out the bifurcation line. Finally there is a certain 
crossover behavior between regions (I and II).

A simple physical realization of a PES such as given by Eq. (\ref{b2}) 
would be a molecule with two
equivalent
coupled wide amplitude coordinates $x$ and $y$ describing, 
for example, proton tunneling (in monograph
\cite{BM}
many such examples are discussed). 
When the coupling between the tunneling particles is neglected, 
the dimensionless PES is represented as a sum of two strongly fluctuating motions:
\begin{eqnarray} 
\label{b3} 
V_0(x , y) = \frac{1}{2}(1-x^2)^2 + \frac{1}{2}(1-y^2)^2 \, .
\end{eqnarray}                                                                                                  
Taking into account symmetric coupling, a term:
\begin{eqnarray} 
\label{b4} 
V_1(x , y) = c(x^2 -y^2)^2 \, ,
\end{eqnarray}  
must be added to Eq. (\ref{b3}). 
After a transformation to coordinates $X=(x+y)/2$ and $Y=(x-y)/2\sqrt 3$, one
obtains the PES given by Eq. (\ref{b2}), provided that the parameters 
$\omega $ and $\alpha $ are related as follows: 
$$ 
\omega = 3\sqrt {2(1-\alpha )} \, .
$$
To explore the behavior of the system with respect to variations of 
parameters $\omega $ and $\alpha $, controlling the shape and the topology 
of the model PES, Eq. (\ref{b2}), the
minimum energy and extreme tunneling trajectories should be fully 
analyzed. Dealing with quantum transitions between different 
states of the system, one must distinguish between adiabatic 
and non-adiabatic transitions. 
Adiabatic transitions take place at $\omega \gg 1$ along the minimum (maximum) energy paths (MEP),
defined by the condition that the PES should have an extremum orthogonal to the MEP:
\begin{eqnarray} 
\label{b5} 
\frac{d Y}{d X} = \frac{\partial V/\partial Y}{\partial V/\partial X} =
\frac{-2(1 -X^2)(1 - \alpha ) +\omega ^2 Y^2}{(X^2 - \alpha )Y^2}
\frac{X}{\omega ^2 Y} \, .
\end{eqnarray}                                                                                                  
From the symmetry of Eq. (\ref{b5}) is clear that $Y^2$ is a function of $1-X^2$, 
and $dY/dX = 0$ on the
surface $X = 0$. 
In the region II of Fig. 1, there exist therefore two types of paths 
satisfying Eq. (\ref{b5}):

(i) two equivalent paths I via the saddle points;

(ii) a shorter path II, $Y = 0$ passing through the maximum.

The extreme tunneling (i.e. instanton) zero energy trajectories 
are described by the equation of motions (''Newton laws'') 
in imaginary time and for the inverted PES (i.e. for $V \to - V$) 
\begin{eqnarray} 
\label{b6} 
{\ddot X} = \frac{\partial V}{\partial X} \, , \, {\ddot Y} = \frac{\partial
V}{\partial Y} \, . 
\end{eqnarray}
At $\alpha   > 0$, 
the trajectories satisfying Eq. (\ref{b6}) lie between the paths 
I and II due to a non-adiabaticity of the transitions caused 
by the finite frequency of the transverse oscillations. 
The instanton action increases when $\alpha $ decreases, 
since the absolute value of the potential at the
saddle points increases as:
\begin{eqnarray} 
\label{b7} 
V^* \equiv V(0 , \pm \sqrt \alpha ) = \frac{1}{2} - \frac{\omega ^2 \alpha
^2}{4(1 -\alpha )} \, . 
\end{eqnarray}
 
\section{Quantum bifurcation.}
In this section, the results of the previous section 
are used to illustrate numerically the appearance of quantum bifurcations 
(already qualitatively discussed in the introduction). 
Standard applications of the instanton method reduce to minimizing 
the non-local action along a one dimensional trajectory. 
Such an approach supposes that there is a single path connecting 
initial and final states. This is evidently not the case for our model. 
As a first step, the equations of motion, Eq. (\ref{b6}), 
are solved and the resulting minimum energy and
extreme tunneling (instanton) trajectories are shown in Fig. 2 for two values 
of the controlling parameter $\alpha $. 
As the instanton trajectories deviate from the minimum energy
paths, the collapse of the two instanton trajectories, when 
$\alpha $ decreases, occurs at a certain finite value
of $\alpha ^* > 0$ which does not coincide with the point 
$\alpha  = 0$ where the PES changes its topology. The
dashed line in Fig. 1 defines region $II^\prime $ within region II, 
where two minimum action trajectories exist, in the remainder 
of region II the minimum action trajectory is the 1D trajectory $Y = 0$. 
Thus, the potential Eq. (\ref{b2}),
admits two types of instantons: a 1D instanton ( 
$Y = 0$ and $\dot Y
= 0$); and 2D instantons where the transverse 
displacements have a maximum at the plane $X = 0$, separating initial and 
final states, while the
transverse
velocities achieve extrema at intermediate points 
where the trajectories deviate from the 1D instanton. 
The contribution of any trajectory to the transition rate 
is a result of the tradeoff between the length of the path and 
the barrier height along this path. Naively, exploiting such 
idea, we come to the bifurcation criterion, which determines the 
critical value of the controlling parameter $\alpha $ 
as the value at which the actions along the both
competing
trajectories become equal. 
We designate the value by $\alpha ^*$. 
However, by definition of instanton actions, the point $\alpha ^*$
corresponds to the harmonic approximation of the potential, 
and therefore this criterion is too crude for our anharmonic 
model potential, Eq. (\ref{b2}). In Fig. 3, the minimum action as 
a function of $\alpha $ is plotted for $\omega = 1$. In
the same figure the action along the minimum energy paths is also represented. 

A more rigorous approach is based on the observation 
that at the bifurcation point a new type of the trajectory (2D) 
appears, and the initial 1D trajectory becomes unstable. 
To study the stability of the trajectory it is sufficient 
to restrict oneself to a Gaussian (harmonic) approximation
for the transverse motion, 
and the instanton approach is therefore adequate. 
The one dimensional instanton solution becomes unstable 
when the lowest eigenvalue of the second functional 
derivative of the action over the trajectory variations becomes zero
$$
\frac{\delta ^2 W}{\delta X^2} = 0 \, .
$$
This lowest eigenvalue is traditionally called 
the stability parameter $\lambda $. When the value of 
$\lambda $ becomes zero (the corresponding value of $\alpha $ will
be called the critical value and denoted $\alpha _c$) 
the tunneling channel centred around an extreme 1D
path
becomes infinitely wide (and the transverse fluctuations 
are no longer Gaussian) and new types of 
the trajectories (2D), possessing a smaller action, appear. 
We believe that this criterium is physical, and it will be shown 
below that just around the value $\alpha _c$ the critical behavior 
of the tunneling splitting takes place
(more precisely the crossover due to quantum zero point fluctuations).

The next step is to determine the wave functions. 
The ground state instanton wave functions can be represented in the form of Eq. (\ref{b1}) as
\begin{eqnarray} 
\label{b8} 
\Psi _0 (X , Y) =
A_0(X , Y) \exp (-\gamma W (X , Y)) \, , 
\end{eqnarray} 
where the functions $A_0(X,Y)$ and $W(X,Y)$ should satisfy the Hamilton-Jacobi and transport equations
\cite{BM}, \cite{BV99} , \cite{BV00} 
\begin{eqnarray}
\label{b9} 
\frac{1}{2}\left (\frac{\partial W}{\partial X}\right )^2 + \frac{1}{2}\left (\frac{\partial
W}{\partial Y}\right )^2 = V(X , Y) \, , 
\end{eqnarray} 
and 
\begin{eqnarray} 
\label{b10} 
\frac{\partial W }{\partial X}\frac{\partial A_0}{\partial X} +
\frac{\partial W }{\partial Y}\frac{\partial A_0}{\partial Y} + \frac{1}{2}\left (\frac{\partial ^2
W}{\partial X^2} + \frac{\partial ^2 W}{\partial Y^2} - 2 E_0 \right )A_0 = 0 \, , 
\end{eqnarray}
$E_0$ is the ground state energy.                                                                         

In Fig. 4, the ground state wave function at the surface $X = 0$ 
is shown for different values of the
parameter $\alpha $. 
Far away from the critical region, the wave functions are localized at the saddle
points while the probability distribution is smeared out 
over the whole region between the potential minima in 
the critical region. The tunneling splitting is given by the expression:
\begin{eqnarray} 
\label{b11} 
\Delta _0 = \int _{-\infty }^{+\infty } \Psi (0 , Y) \left (\frac{\partial
\Psi }{\partial Y} \right ) _{X=0} d Y \, . 
\end{eqnarray} 
The value of this splitting given by the 
probability flow integrated over the dividing plane $X=0$ depends on
the exponential factor, 
$\exp(-\gamma W^*)$, where $W^*$ is Euclidean action between the minima.               
Appart from this leading term, $\Delta $ depends on
prefactors characterizing both type trajectories. In order to eliminate the leading
term contribution and visualize the critical behavior of $\Delta $, $\Delta _0
\exp(\gamma W^*)$ is plotted as a function of $\alpha $ in Fig. 5.                                                  
The width of the maximum around the critical value, $\alpha = \alpha _c$,
is determined by quantum (zero point) fluctuations.

Tunneling leads to an energy splitting 
(tunneling splitting, $\Delta _0$) between two otherwise 
degenerate states and the eigenstates are
symmetric and
antisymmetric combination of these states. 
This tunneling splitting between the two lowest eigenstates of 
the Hamiltonian is therefore the quantity of interest. 
Alternatively, the rate of oscillation between the 
two classically degenerate states can be measured after the 
system has been prepared in one of these states. 

At small values of $\alpha $ some new phenomenon could occur 
due to these quantum fluctuations. In this
region,
at the surface $X = 0$, the depth of the wells becomes 
so small that wave functions are effectively
delocalized due to quantum fluctuations. 
The natural (though qualitative) criterion of 
the delocalization should compare amplitude of zero point oscillations along $Y$ 
in one well with the
distance between wells. 
The same criterion as given above, could be 
formulated as the condition that there are no discrete 
energy levels in a shallow well. Both criteria compare 
the energy difference between the maximum and 
the saddle point of the PES, Eq. (\ref{b2}), $V(0,0) - V(0 , \sqrt \alpha )$, 
with the energy of
zero-point oscillations in
one-well, $\tilde \omega /\gamma $, where $\tilde \omega $  
is the frequency of oscillations around $\pm \sqrt \alpha $ in
the potential $V(0,Y)$,
renormalized by anharmonic corrections. 
Taking into account all numerical factors we obtain:
\begin{eqnarray} 
\label{b12} 
\frac{\alpha _c^2 \omega ^2}{4(1 - \alpha _c)} \simeq (1 + \sqrt
2)\frac{\omega }{\gamma } \sqrt {\frac{2 \alpha _c}{1-\alpha _c}} \, ,
\end{eqnarray}                                                                                                  
Eq. (\ref{b12}) is fulfilled for $\alpha _c = 0.56$ (at $\omega  = 1$) 
as found in our numerical
computations shown in Fig. 5. 

The square root divergence of the localization length at 
$\alpha _c$ corresponds to the critical behavior 
of the correlation length in standard classical second
order
phase transitions. 
The tunneling splitting $\Delta _0$, calculated above, is our main result. 
$\Delta _0$ is the product of an
exponential and
pre-exponential factor, and only the pre-factor demonstrates 
the critical behavior (more precisely - the crossover behavior) 
shown in the Fig. 5. It may, of course, not be easy 
(if possible at all) to measure directly the contribution 
of a small pre-factor with a singular behavior as a function of $\alpha $ when 
$\Delta _0$ is dominated by the
large exponential factor with a regular behavior. 
Given that a measurement of a non-linear susceptibility of 
the system is equivalent to the measurement of the corresponding derivative 
of $\Delta _0$ over the controlling parameter $\alpha $ 
offers one possibility. For example, when a local
dipole moment is
associated with the tunneling particles in a system 
with inversion symmetry, an applied external electric 
field will make the two 2D paths inequivalent, 
while the single 1D path would be unaffected. 
Let us emphasize that the symmetry breaking phenomenon 
takes place due to tunneling processes only and 
is thus by its nature a dynamic phenomenon. 
We will discuss this issue in the next section. 

The second possibility to observe this kind of critical 
(crossover) behavior is based on the general Herring-Lifshits formula 
(see Eq. (\ref{b11}) and  \cite{LL65}), 
which shows that the splitting $\Delta _0$ is determined 
by the integral over the $Y$
coordinate at the dividing surface $X = 0$, 
and is thus proportional to the total width of the tunneling
channels. Therefore, the effect we got, might be considerably 
enhanced in 3D systems due to increase of 
the phase volume of tunneling channels. A 3D model PES, of which 
our 2D model PES, Eq. (\ref{b2}), is a plane section, is given by $V(X,\rho )$, where ($\rho  = 
(Y^2 + Z^2)^{1/2}$.

\section{Anisotropy.} 
\label{anis}

The PES, investigated in the previous sections, 
is symmetric with respect to the reflection $Y \to - Y$. 
Both tunneling particles are therefore identical
as are the two tunneling paths. 

As mentioned above, breaking this symmetry may be useful, 
in order to expose more clearly the singular behavior of the 
pre-factor. Indeed, the 
response should to expose a qualitatively 
different behavior with respect to any factor 
breaking the symmetry near the critical region 
(i.e. close to the bifurcation point) or far from it. 
This idea, the so-called fluctuation - dissipation theorem 
relating correlation and response functions, is well known 
for conventional thermodynamic phase transitions. 
Two mechanisms of breaking this mirror symmetry, 
which are easily realized, will be 
analyzed below. First, partial 
deuteration will make the tunneling particles different. 
Second, as mentioned above, an external electric field 
may render inequivalent the tunneling 
channels for the two particles. Both mechanisms 
can be described in a phenomenological way by adding 
the following anisotropic perturbation to the bare symmetric PES, Eq. (\ref{b2}): 
\begin{eqnarray} 
\label{b13} V_a(X , Y) = \beta (1 - X^2) Y \, ,
\end{eqnarray}                                                                                                  
The parameter 
$\beta $ describes the strength of the anisotropy.
In the potential (\ref{b13}) 
higher order anisotropic terms, $Y^3(1-X^2)$, and so on,
were neglected for the sake of simplicity.
Even though, strictly speaking, this assumption is not justified,
since $Y$ is a strongly fluctuating degree of
freedom,
this approximation correctly identifies the characteristic scales in the
problem and all qualitative features of its behavior. 
For hydrogen - deuterium substitution of one of the
two tunneling atoms, for example, the resulting anisotropy parameter is easily evaluated as: 
\begin{eqnarray} 
\label{b14} 
\beta = \frac{1}{4} \left (\sqrt {\frac{m_1}{m_0}} - 1 \right ) \, ,
\end{eqnarray}                                                                                                  
where $m_0$ and $m_1$ are masses of the $H$ 
and $D$ isotopes. In an applied electric field, the anisotropy
parameter $\beta $ will be proportional to the field strength. 
The precise choice of the anisotropic PES,
Eq. (\ref{b13}), is delicate, and depends on the detailed structure of the system under
consideration. Aiming at
a qualitative description, we have chosen the simplest form of 
breaking the $Y \to - Y$ symmetry while the symmetry $X \to - X$ is preserved, 
so that the two potential
well are still equivalent.

The analysis of this modified PES is analogous 
the one presented in the section II for the symmetric PES. 
In the $\alpha \, , \, \beta $ parameter phase plane, 
we should first find all minimum energy and
extreme tunneling
trajectories. There are 4 different regions on this plane as shown in Fig. 6. 
In the region II one maximum and two saddle points exist, in region I 
there is only one saddle point. The behavior is more subtle 
in region III, where the saddle point that is more distant 
from the maximum is transformed into the minimum (i.e. the $X$ -''oscillation'' 
frequency changes its
sign). By the dashed lines we depicted also the sub-region $II^\prime $, where there are two extreme
tunneling trajectories. In the remaining part of the region II, 
corresponding to relatively small values of $\alpha $, 
the extreme tunneling trajectory close to the
maximum becomes unstable.

In Fig. 7 we show the minimum energy and extreme tunneling 
(instanton) trajectories for a small value of the anisotropy 
parameter ($\beta  = 0.05$) and for different values of $\alpha $. 
It illustrates that even a small
anisotropy lifts the
degeneracy of two instanton trajectories. In addition, the straight 
line path II, passing through the maximum $\{X = 0\, , \,  Y = 0\}$, 
for the symmetric PES, Eq. (\ref{b2}),
becomes curvilinear and deviates from $Y = 0$. Hence for the anisotropic 
PES ($\beta \neq  0$) all regions in the
phase diagram with different types of the trajectories 
have the same symmetry, and therefore only first 
order bifurcations between the regions are allowed. 
Finally, in Fig. 8, the actions along the instanton 
and the minimum energy paths are plotted. The 
figure shows again that the anisotropy, Eq. (\ref{b13}), 
removes the degeneracy, and that two instanton
trajectories appear when $\alpha $ exceeds a certain threshold value. 
This disappearance of one instanton
trajectory is a specific feature of 
discontinuous first order bifurcations. 
Concluding this section, we emphasize that it is 
this sensitivity to anisotropy, which can lead to a drastic change of behavior.

\section{Experimental consequences} 
\label{exp}

Bifurcations of minimum energy paths (due to 
the presence of more than one saddle point separating stable configurations of 
the PES) are rather common in molecules 
with several strongly fluctuating coordinates. 
In non-rigid molecules, dynamically strongly 
fluctuating coordinates of this sort of are typically 
different combinations of hydrogen transfer, 
hindered rotation of $-XH_n$ groups, or inversions. 
Molecular systems with two hydrogen transfers
(synchronous or stepwise) attract special attention, as 
these processes are thought to be relevant 
for many biological processes, including so-called tautomeric reactions 
\cite{be1}, \cite{be2}, \cite{be3}, \cite{BM}.  

Two proton exchange in pairs of $OH \cdots O$ fragments of various carbonic acid dimers is an example of
synchronous tunneling. In our model, this transfer corresponds to a one dimensional trajectory in region
II, and the longitudinal $X$ and transverse $Y$ coordinates are symmetric and antisymmetric combinations
of proton displacements $d_1$, $d_2$: 
\begin{eqnarray} 
\label{b15} 
X = \frac{1}{2}(d_1 + d_2) \, ; \, Y =
\frac{1}{2\sqrt 3}(d_1 - d_2) \, . 
\end{eqnarray} The coefficients are chosen such that Eq. (\ref{b2}) is
the sum of Eqs.  (\ref{b3}) and (\ref{b4}), and the results of Sections II and III can therefore be used.
For the 1D path $Y = 0$, with one saddle point $\{X = 0\, , \, Y = 0\}$, the displacements of the both
tunneling protons are always (i.e. at any point of the trajectory) equal to each other ($d_1 = d_2$).  
In
the pure classical limit, the width of the tunneling channel is determined by the corresponding potential
curvature:  
$$ 
\Delta _\perp = \left (\frac{\partial ^2 V}{\partial Y^2}\right )_{0 , 0}^{1/2} \, ,
$$                                                                                                              
Within the harmonic approximation, there is a mean field ($\propto \alpha ^{-1/2}$) 
divergency. In the
plane
$X = 0$, the anharmonic $Y^4$ contribution provides a cutting-off of this divergency. 

Quantum, zero point
transverse fluctuations spread out this singularity, 
and the width of the tunneling channel depends on
$\alpha $ even in region II of Fig. 1. The characteristic potential barrier height, $V^*$, 
for the
synchronic two
proton transfer in the carbonic acid dimers is $V^* \simeq  (10 - 15)$ Kcal/mol 
(5000 - 7000 K), while the
energy $E_{st}$ corresponding to a step wise transfer 
(that is paths $d_1 = 0\, , \, d_2 = 1$ or $d_1 = 1\, , \,  d_2 = 0$) is
two or even three times larger 
\cite{BM88}, \cite{LK97}.  
On the other hand, within the frame of our model PES, the 
ratio $E_{st}/V^*$ depends
on the controlling parameter $\alpha $ as follows: 
\begin{eqnarray} 
\label{b16} \frac{E_{st}}{V^*} = 1 - \frac{3}{2}\alpha \, , \, \, \alpha < 0 \, .
\end{eqnarray}                                                                                                  
Inserting the numerical values of the characteristic energies given above into
Eq. (\ref{b16}), one can see that even in 
the systems where the strongly fluctuating motions are believed
to be
strongly correlated, the controlling parameter $\alpha $
is close to -0.5, and quantum fluctuations must be
taken into account to describe correctly the tunneling dynamics. 

Many examples of intermediate (between
synchronic and step wise) dynamics exist also, 
one example being the so-called free base porphyrin
compounds \cite{BM88}, \cite{SM89}. 
In these compounds with molecular symmetry $D_{4h}$, 
four nitrogen atoms (numbered
clockwise: $A$, $B$, $C$, $D$) form a square and two mobile 
protons $a$ and $b$ occupy positions such that
configurations ($aA$, $bC$) and ($aB$, $bD$) are equivalent 
minima, while configurations ($aA$, $bD$) and ($aB$,
$bC$) are
equivalent saddle points (taking into account clockwise 
motions only). Our model PES, Eq. (\ref{b2}), can be
adapted to this case with . According to experimental 
data and quantum chemistry calculations  \cite{BM88},
\cite{SM89} 
the
energy of the saddle points is of the order 
of 0.3 -0.5 of the energy at the maximum ($X = 0\, , \, Y = 0$). 
Thus
from Eq. (\ref{b7}) we come 
to the estimation $\alpha \simeq 0.3 - 0.4$. 
For these intermediate values of $\alpha $, the
quantum (minimum action) and the classical (minimum potential) 
trajectories are quiet different. Indeed, the
classical and tunneling trajectories pass through 
the saddle points, and the maximum, respectively. The
existence of saddle points lead to a 
considerable broadening of the tunneling channel as it is illustrated
in the Fig. 3. Similar 
considerations are easily applied to the hydrazine molecule with two coupled
inversion motions (see \cite{LK97} 
 and references therein). 

For the partially deuterated molecule, we can also
evaluate the anisotropy parameter $\beta $ entering the PES, Eq. (\ref{b12}). 
According to Eq. (\ref{b13}) $\beta \simeq 0.1$, and
the $H - D$
isotopomer remains therefore in the same region 
I of the phase diagrams as protonated porphyrin. The main
conclusion of this short analysis is that 
our simple model gives a fairly realistic representation of
non-rigid molecules. 

We have shown that 
effects due to competition of trajectories and quantum
fluctuations are relevant in a broad range 
of the parameters entering our model. The examples discussed
above lead to the conclusion that the widely accepted 
classification of synchronic and step wise motions
must be used with care, since in typical cases 
both types of motions are involved. This implies also that
the traditional \cite{CO}  
Gaussian approximation 
(small fluctuations around one extreme tunneling trajectory)
in not valid in the case of strong coupling 
between both tunneling particles, and one must take into
account both competing tunneling channels. Similar 
phenomena might play a role in pairing of isolated
nucleic-acid bases in the absence of the DNA backbone. 
In Ref. \cite{NK00}   
spectroscopic characteristics of the
hydrogen bonding in isolated guanine-cytosine ($G - C$) 
and guanine-guanine ($G - G$) base pairs have been
investigated. The results show that the gas phase $G - C$ 
base pair adopts a single configuration, whereas
$G - G$
exists in two different configurations.  
We already mentioned in the section II that the effect, obtained
here, may be considerably enhanced in 3D systems 
due to the increase of the phase volume of the tunneling
channels (our 2D model PES can be considered 
as a section of the corresponding 3D space). Among systems,
where a 3D generalization of our model could be applicable, 
are cubic alkali halide crystals doped with a
light atomic or molecular ions as substitutional impurity. 
Due to mismatch in size, these impurities
frequently occupy off-centre positions in the lattice. 
There is a small number of equivalent off-centre
positions, so that the impurity ground state is 
degenerate, and tunneling transitions between these
positions are observed. However, the vibrational 
frequencies of off-centre impurities are usually much
smaller than characteristic host lattice phonon 
frequencies, and therefore the coupling to the lattice
vibrations is relatively small. As a first approximation, 
one can visualize the system as one particle in
a potential where the tunneling transitions 
between the initial and the final state can proceed in several
alternative ways. In the trajectory language 
it signifies the possibility of competition (and bifurcation)
between extremal paths of different types. 
Since transitions between different off-centre positions are
associated to a charge displacement, the corresponding 
PES changes upon application of external electric
field or mechanical stress. 
We anticipate that the bifurcations described above by our model could be
observable for off-centre impurities as well, 
and that understanding this mechanism will be essential to
predict and to describe the behavior. 
A more specific study of 3D systems undergoing this kind of the
quantum bifurcations, might become appropriate as 
suitable experimental results become available. 

A completely different system might also be 
a physical realization of our model. The tunneling of magnetic
spins has recently received much attention (see e.g. \cite{GB95}) 
in view of its promise as one of the few
realistic candidates for quantum computing. 
Conventional magnetic materials used in the experiments
contain many domains, each possessing its own set 
of parameters. Besides the spins interact with the
crystal matrix and complicates the physical 
picture with respect to our model. However, quantum tunneling
of spin is also possible in a spinor 
condensate trapped in double - well potential \cite{GB95})  
and this system
possesses several decisive advantages 
compared with more conventional solid state materials and is more
suitable for a description in the frame work 
of our model (the system is characterized by a few simple
parameters amenable to experimental control). 

\section{Conclusion.}
\label{conc}                                                                                                    

We have presented here a path-integral
approach to treat a 2D model of quantum bifurcations. 
A PES, the shape of which is determined by a
continuous parameter, offers a natural way to examine 
quantum instability phenomena. Candidates that
realize our model for quantum bifurcations are 
non-rigid molecules, spin condensates, and some other
systems. In any system exhibiting some kind of 
quantum phase transitions it is important to understand how
its genuine quantum aspects evolve throughout the 
transition. We investigated the behavior of the ground
state wavefunction undergoing qualitative 
changes at a quantum instability, but other indicators of the
phase transition could be examined in the 
same way (in a very recent publication, for example, the authors
studied wave function entanglement phenomena \cite{OA02}). 

A second order bifurcation takes place for a 2D model
PES with mirror symmetry, which changes its topology at a certain 
critical value $\alpha _c$ of the continuous
parameter $\alpha $. At $|\alpha - \alpha _c| \gg \alpha _c$ 
only one minimum action trajectory is
essential for semiclassical
description of the particle motion, while in the ''critical'' 
region at $|\alpha - \alpha _c| \ll \alpha _c$ the behavior is
governed by two different trajectories almost degenerate with 
respect to the magnitude of the action. In
our model, the competition between trajectories 
plays the role of competing interactions in many body
systems experiencing quantum phase transitions. 
The divergence in correlation length is mapped onto a
singular behavior of the localization length. 
Due to zero point quantum fluctuations, the system manifests
a smoothed crossover behavior only instead 
of a sharp bifurcation. When the PES becomes asymmetric, a
first order phase transition is 
associated with the disappearance (in a certain region of the potential
parameters) of one minimum action trajectory. 
We illustrate the results by numerically investigating this
behavior. The approach is general and has potential 
applicability for large (many particle) systems. 
Furthermore, it is possible to fabricate structures which are small
enough (mesoscopic) for the electronic transport to be largely coherent,
so that the wave (quantum) properties of electrons dominate.
Note also that the usefulness of our results can be substantially
enhanced by the advent of semiconductor heterostructures in which
the potential for the electrons can be tuned and varied
in space with great precision.

One
more an interesting line of thought for 
future work would be to analyze, in analogy with the approach
presented here, a phase stability with 
respect to small fluctuations for conventional phase transitions.
It is known for Landau-Ginzburg type 
models \cite{stat}, that the corresponding 
Landau-Khalatnikov equations have
a form analogous to the Schr\"odinger equation 
(where a coordinate plays the role of an imaginary time)
\cite{AF91}.  

Our results can in principle be tested also by investigations 
of low temperature properties of
disordered materials, containing certain point 
defects that undergo atomic tunneling \cite{ES98}. 
Recently \cite{KF99}, \cite{WU02} 
the low temperature dielectric properties 
of certain multi-component glasses were tentatively
assigned to the existence of coupled two-level tunneling 
systems. This means that several tunneling paths
between potential minima exist, as in the model presented here. 

It is important to notice that the
investigation of our paper refers to a singular behavior 
in the ground state of the system, as, strictly
speaking, quantum phase transitions occur only at $T = 0$. 
Because all experiments are necessarily done at
some nonzero temperature, care must be taken 
when comparing our theoretical results to experiment, and we
have to keep in mind the consequences of the $T = 0$, and 
$\alpha = \alpha _c$ singularity on physical properties at $T > 0$. 
Technically, finite temperature
effects also can be incorporated into the instanton method. 
To do this,
one should take into account the finite 
period ($1/T$) of instanton trajectories (an analogous approach was
developed for the WKB method \cite{WL85}, \cite{GO87}). 
When these contributions become comparable to zero point quantum
oscillations, the quantum phase transition is smeared out.

\acknowledgements 
The research described in this publication was made possible in part by RFFR 
Grants. 
We thank A. H\"uller and P. Nozi\`eres and J.Lajzerowicz for thought-provoking discussions.
One of us (E.K.) is indebted to INTAS Grant (under No. 01-0105) for partial support. 
                                                                                                              
\newpage

\centerline{Figure Captions.}
 
Fig. 1                                                                                                          

$\omega ^2 \, - \, \alpha $ phase diagram. 
In the region II the PES has two minima and two saddle
points, only in the region $II^\prime $ two
minimum action paths coexist with 1d trajectory $Y = 0$, 
in the region I the PES has two minima and one
saddle point and in the region III more than two minima.

Fig. 2 

Equipotential map for 2D PES Eq. (\ref{b2}); 
dashed lines designate minimum energy trajectories, and the solid
lines show the minimum action (tunneling) trajectories. 
(a) $\omega  = 1\, , \, \alpha = 0.55$ , 
(b) $\omega  = 1\, , \, \alpha = -0.20$ .

Fig. 3 

The action $W^*$ as a function of the 
parameter $\alpha $ at $\omega  = 1$ for: 
(1) the minimum action tunneling trajectory ,
(2) the 1d path, $Y = 0$ ,  and (3) the minimum energy trajectory.

Fig. 4 

Ground state wave functions ($n = 0$)
at $X = 0$ for $\omega  = 1$ and $\gamma = 20$: 
(1) $\alpha = 0.625$ ; (2) $\alpha = 0.55$ ; (3) $\alpha = 0.3$.

Fig. 5 

Tunneling splitting as a function 
of $\alpha $ for $\omega  = 1$, and $\gamma = 20$ 
in the critical region (the dashed line is a
continuation to the stability region).

Fig. 6 

The phase diagram for the anisotropic PES, Eq. (\ref{b13}),
in the $\alpha - \beta $-plane with 
$\omega  = 1$. In the region II the the PES has one maximum
and two saddle points and in the region I only one saddle point. 
In the region III one of the saddle points from region 
II (more distant from the maximum) is transformed into the minimum. 
In the region $II^\prime $ there are two minimum action trajectories.

Fig. 7 

Equipotential map for the anisotropic PES. 
Instanton (solid lines) and minimum energy 
(dashed lines) trajectories. 
$\omega =  1$, $\beta  = 0.05$: 
(a) $\alpha = 0.58$; (b) $\alpha = 0.45$; (c) $\alpha = 0.20$.

Fig. 8 
The action $W^*$ as a function 
of $\alpha $ along the instanton (solid lines) 
and minimum energy (dashed
lines) paths. 
$\omega  = 1$, and $\beta  = 0.05$. 
The curves 1 and $1^\prime $ correspond to the trajectories more distant from
the maximum, whereas lines 2 and $2^\prime $ correspond to the trajectories close to the maximum.

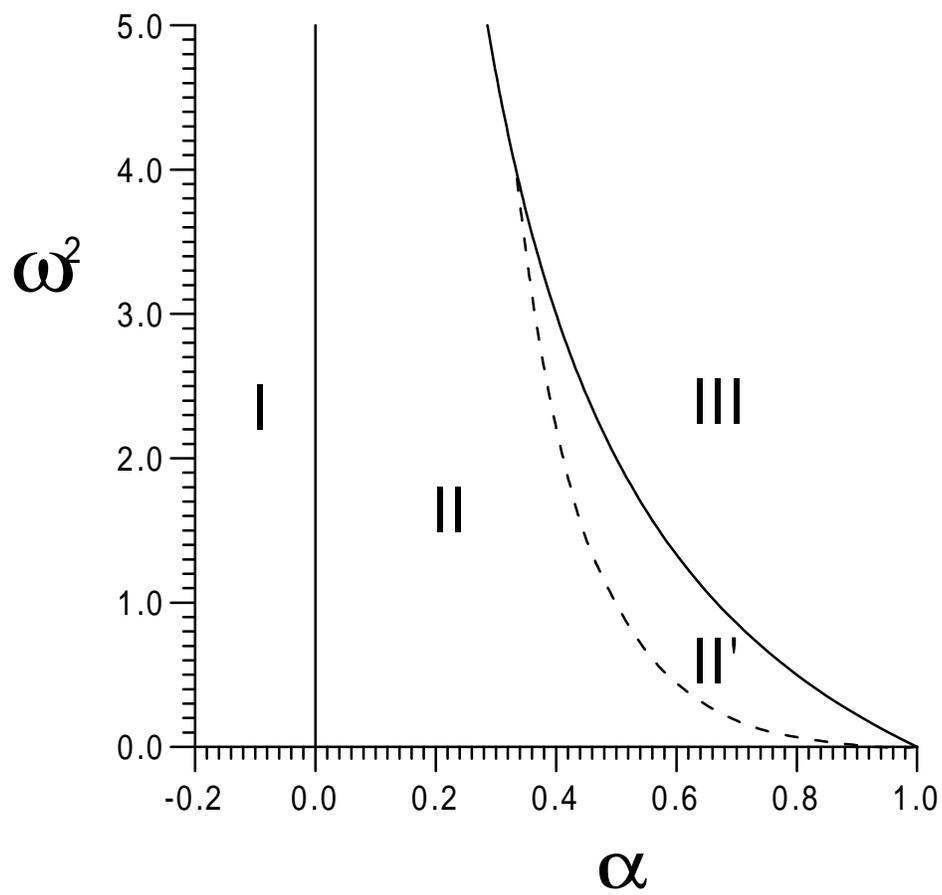
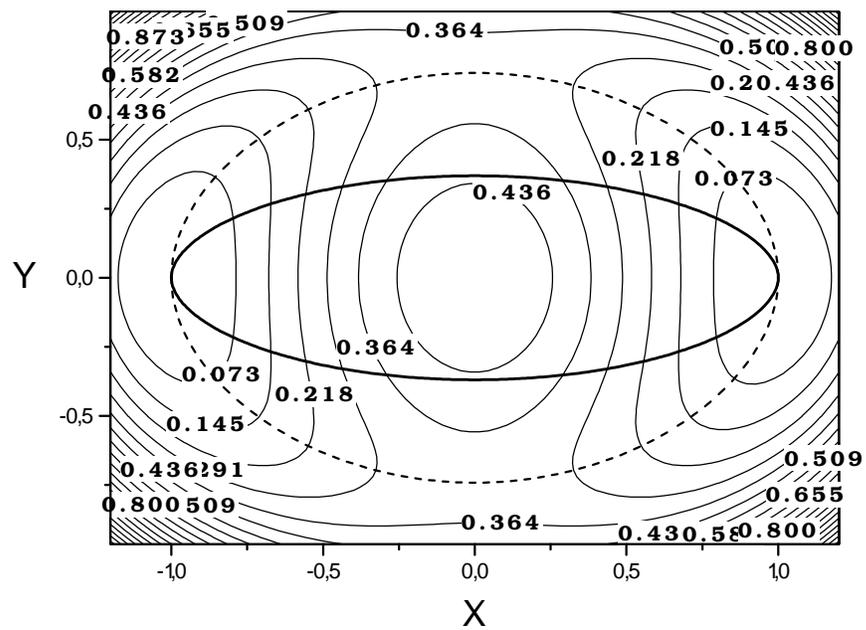
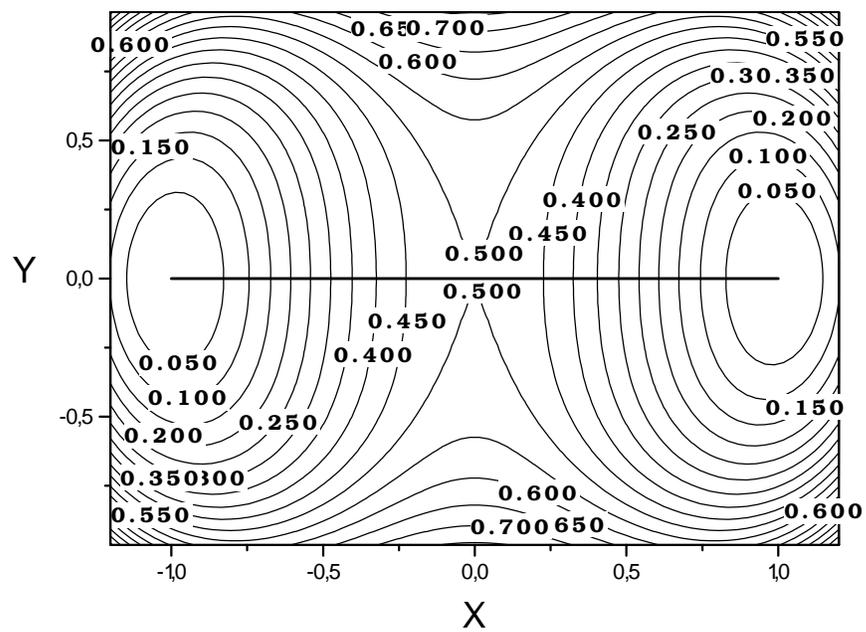
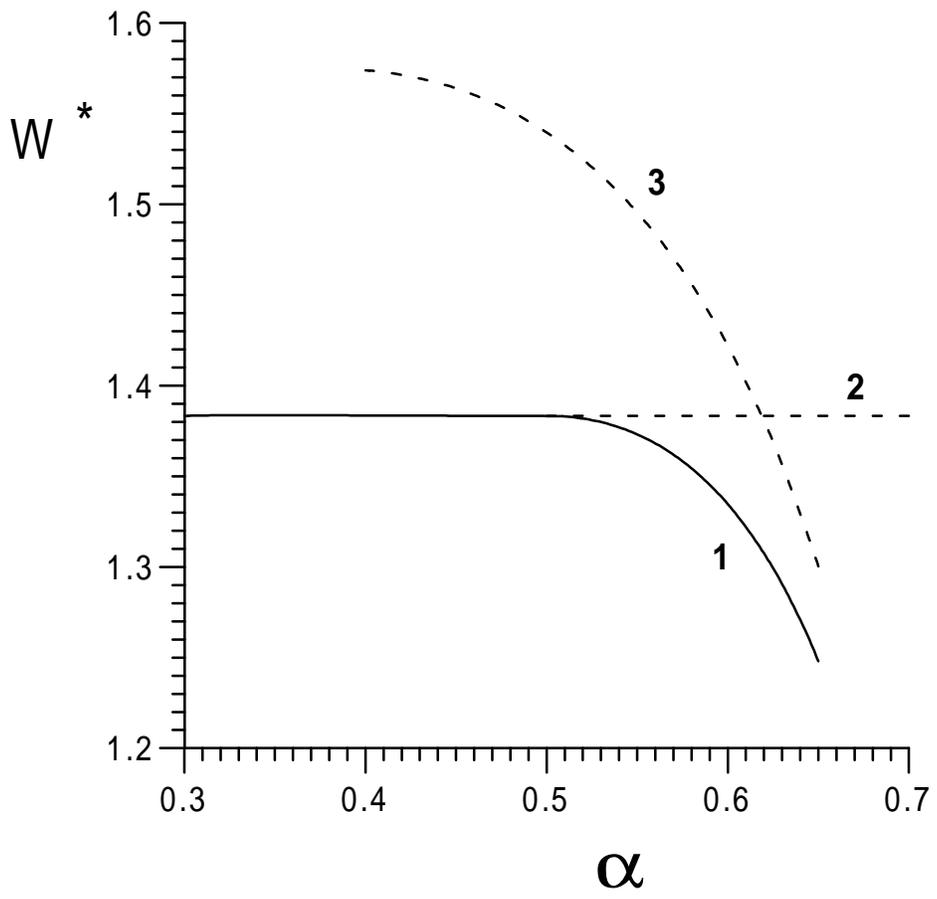
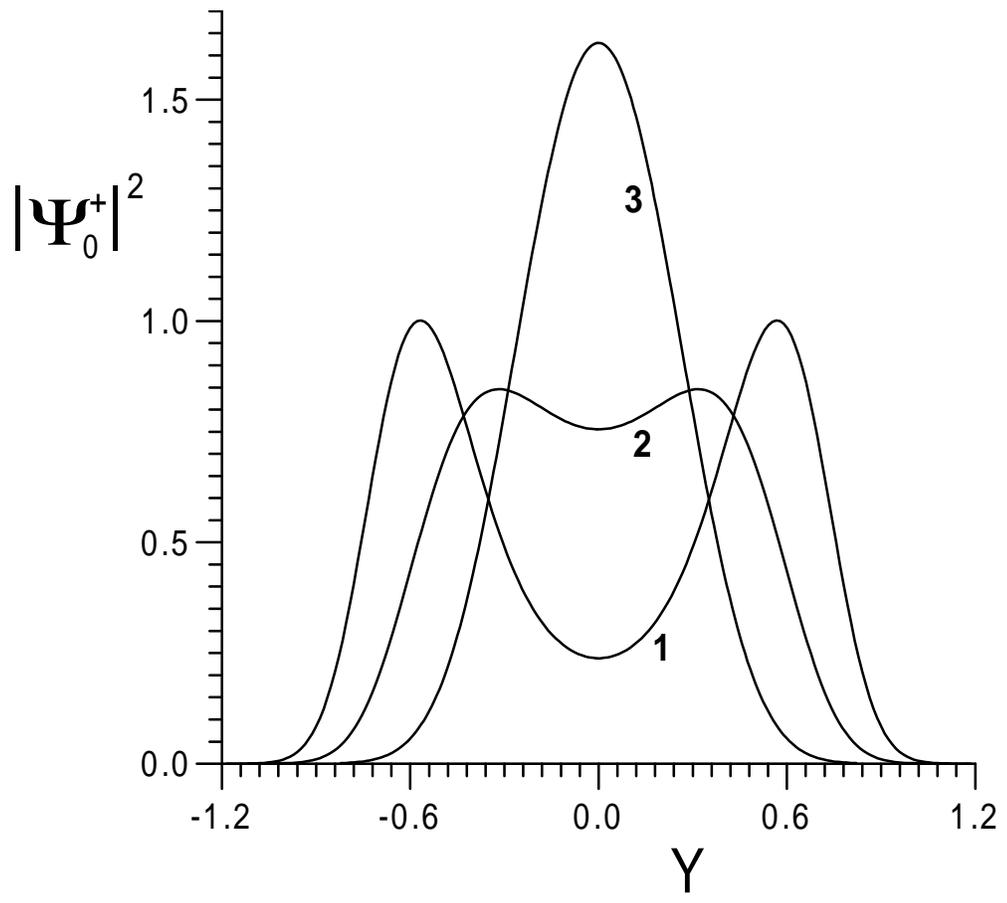
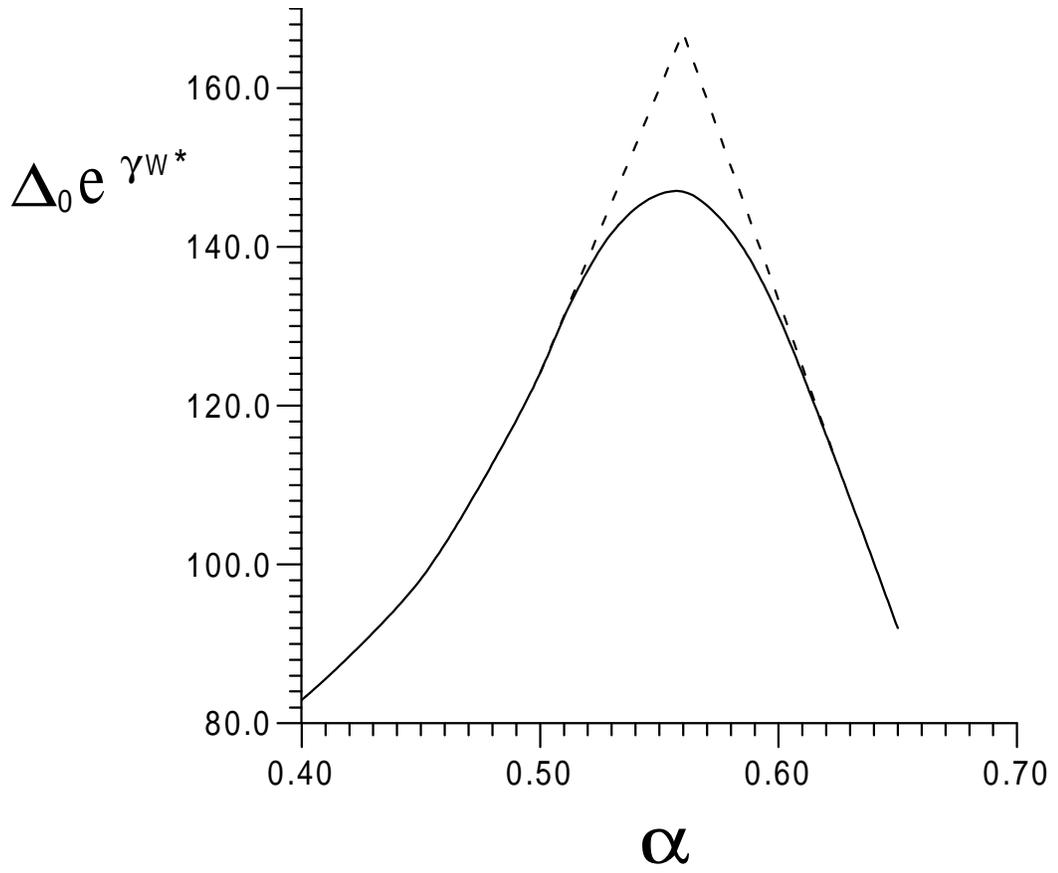
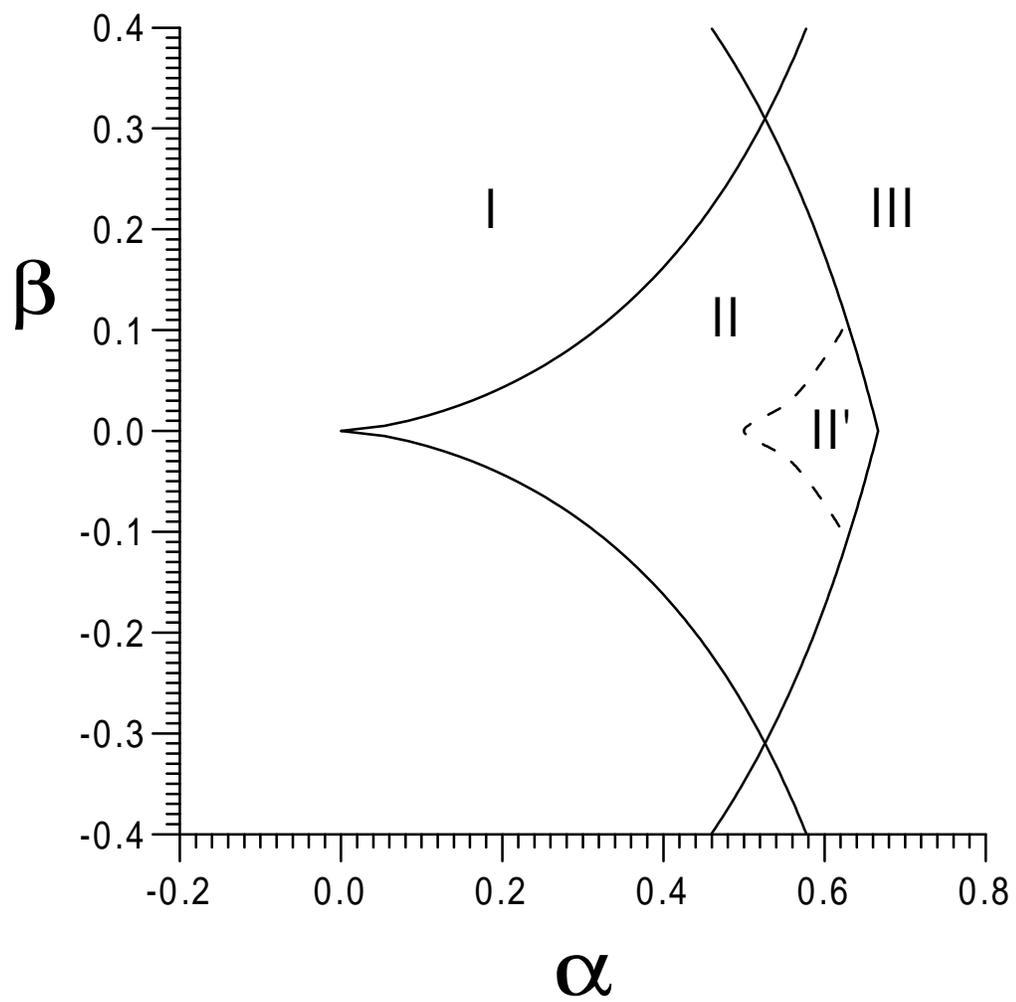
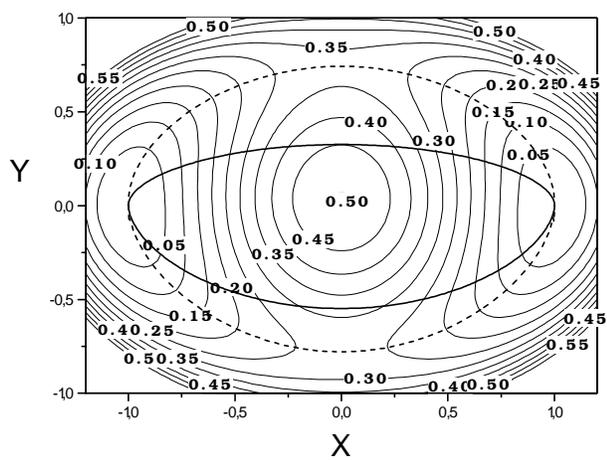
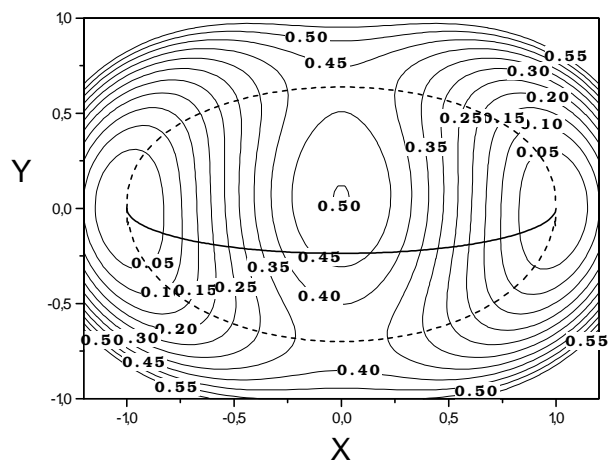
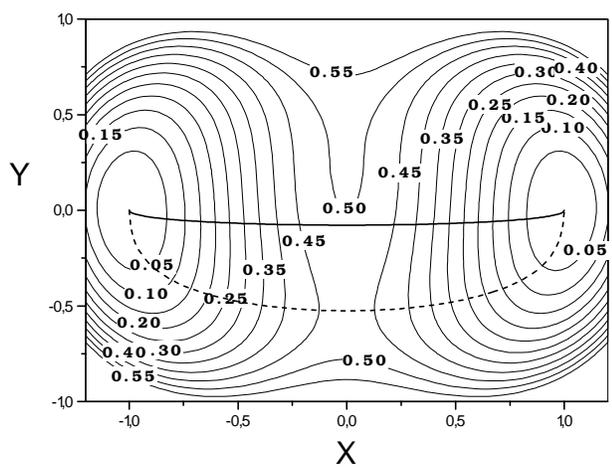
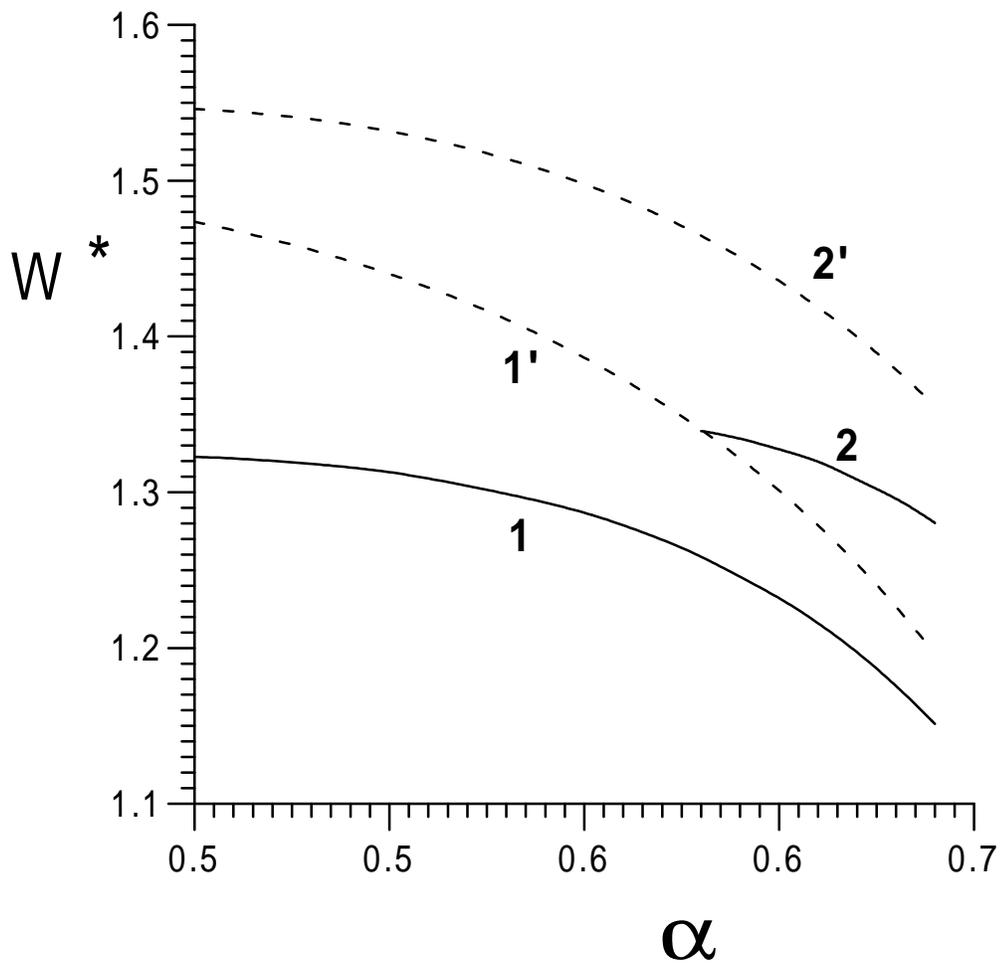
\end{document}